\documentstyle[twoside,psfig]{article}
\catcode`\@=11
\long\def\@makefntext#1{
\protect\noindent \hbox to 3.2pt {\hskip-.9pt  
$^{{\eightrm\@thefnmark}}$\hfil}#1\hfill}		

\def\@makefnmark{\hbox to 0pt{$^{\@thefnmark}$\hss}}	
	
\def\ps@myheadings{\let\@mkboth\@gobbletwo
\def\@oddhead{\hbox{}
\rightmark\hfil\eightrm\thepage}   
\def\@oddfoot{}\def\@evenhead{\eightrm\thepage\hfil
\leftmark\hbox{}}\def\@evenfoot{}
\def\sectionmark##1{}\def\subsectionmark##1{}}

\oddsidemargin=\evensidemargin
\newcounter{sectionc}\newcounter{subsectionc}\newcounter{subsubsectionc}
\renewcommand{\section}[1] {\vspace{12pt}\addtocounter{sectionc}{1} 
\setcounter{subsectionc}{0}\setcounter{subsubsectionc}{0}\noindent 
	{\tenbf\thesectionc. #1}\par\vspace{5pt}}
\renewcommand{\subsection}[1] {\vspace{12pt}\addtocounter{subsectionc}{1} 
	\setcounter{subsubsectionc}{0}\noindent 
	{\bf\thesectionc.\thesubsectionc. {\kern1pt \bfit #1}}\par\vspace{5pt}}
\renewcommand{\subsubsection}[1] {\vspace{12pt}\addtocounter{subsubsectionc}{1}
	\noindent{\tenrm\thesectionc.\thesubsectionc.\thesubsubsectionc.
	{\kern1pt \tenit #1}}\par\vspace{5pt}}
\newcommand{\nonumsection}[1] {\vspace{12pt}\noindent{\tenbf #1}
	\par\vspace{5pt}}

\newcounter{appendixc}
\newcounter{subappendixc}[appendixc]
\newcounter{subsubappendixc}[subappendixc]
\renewcommand{\thesubappendixc}{\Alph{appendixc}.\arabic{subappendixc}}
\renewcommand{\thesubsubappendixc}
	{\Alph{appendixc}.\arabic{subappendixc}.\arabic{subsubappendixc}}

\renewcommand{\appendix}[1] {\vspace{12pt}
        \refstepcounter{appendixc}
        \setcounter{figure}{0}
        \setcounter{table}{0}
        \setcounter{lemma}{0}
        \setcounter{theorem}{0}
        \setcounter{corollary}{0}
        \setcounter{definition}{0}
        \setcounter{equation}{0}
        \renewcommand{\thefigure}{\Alph{appendixc}.\arabic{figure}}
        \renewcommand{\thetable}{\Alph{appendixc}.\arabic{table}}
        \renewcommand{\theappendixc}{\Alph{appendixc}}
        \renewcommand{\thelemma}{\Alph{appendixc}.\arabic{lemma}}
        \renewcommand{\thetheorem}{\Alph{appendixc}.\arabic{theorem}}
        \renewcommand{\thedefinition}{\Alph{appendixc}.\arabic{definition}}
        \renewcommand{\thecorollary}{\Alph{appendixc}.\arabic{corollary}}
        \renewcommand{\theequation}{\Alph{appendixc}.\arabic{equation}}
        \noindent{\tenbf Appendix \theappendixc #1}\par\vspace{5pt}}
\newcommand{\subappendix}[1] {\vspace{12pt}
        \refstepcounter{subappendixc}
        \noindent{\bf Appendix \thesubappendixc. {\kern1pt \bfit #1}}
	\par\vspace{5pt}}
\newcommand{\subsubappendix}[1] {\vspace{12pt}
        \refstepcounter{subsubappendixc}
        \noindent{\rm Appendix \thesubsubappendixc. {\kern1pt \tenit #1}}
	\par\vspace{5pt}}

\newcommand{\textlineskip}{\baselineskip=13pt}
\newcommand{\smalllineskip}{\baselineskip=10pt}

\def\eightcirc{
\begin{picture}(0,0)
\put(4.4,1.8){\circle{6.5}}
\end{picture}}
\def\eightcopyright{\eightcirc\kern2.7pt\hbox{\eightrm c}} 

\newcommand{\copyrightheading}[1]
	{\vspace*{-2.5cm}\smalllineskip{\flushleft
	{\footnotesize Modern Physics Letters A #1}\\
	{\footnotesize $\eightcopyright$\, World Scientific Publishing
	 Company}\\
	 }}

\newcommand{\publisher}[2]{{\begin{center}\footnotesize\smalllineskip 
	Received #1\\
	Revised #2
	\end{center}
	}}

\newcommand{\bibit}{\nineit}
\newcommand{\bibbf}{\ninebf}
\renewenvironment{thebibliography}[1]
	{\frenchspacing
	 \ninerm\baselineskip=11pt
	 \begin{list}{\arabic{enumi}.}
        {\usecounter{enumi}\setlength{\parsep}{0pt}     
	 \setlength{\leftmargin 12.7pt}{\rightmargin 0pt} 
         \setlength{\itemsep}{0pt} \settowidth
	{\labelwidth}{#1.}\sloppy}}{\end{list}}

\newcounter{itemlistc}
\newcounter{romanlistc}
\newcounter{alphlistc}
\newcounter{arabiclistc}

\newcommand{\fcaption}[1]{
        \refstepcounter{figure}
        \setbox\@tempboxa = \hbox{\footnotesize Fig.~\thefigure. #1}
        \ifdim \wd\@tempboxa > 5in
           {\begin{center}
        \parbox{5in}{\footnotesize\smalllineskip Fig.~\thefigure. #1}
            \end{center}}
        \else
             {\begin{center}
             {\footnotesize Fig.~\thefigure. #1}
              \end{center}}
        \fi}

\newcommand{\tcaption}[1]{
        \refstepcounter{table}
        \setbox\@tempboxa = \hbox{\footnotesize Table~\thetable. #1}
        \ifdim \wd\@tempboxa > 5in
           {\begin{center}
        \parbox{5in}{\footnotesize\smalllineskip Table~\thetable. #1}
            \end{center}}
        \else
             {\begin{center}
             {\footnotesize Table~\thetable. #1}
              \end{center}}
        \fi}

\def\@citex[#1]#2{\if@filesw\immediate\write\@auxout
	{\string\citation{#2}}\fi
\def\@citea{}\@cite{\@for\@citeb:=#2\do
	{\@citea\def\@citea{,}\@ifundefined
	{b@\@citeb}{{\bf ?}\@warning
	{Citation `\@citeb' on page \thepage \space undefined}}
	{\csname b@\@citeb\endcsname}}}{#1}}

\newif\if@cghi
\def\cite{\@cghitrue\@ifnextchar [{\@tempswatrue
	\@citex}{\@tempswafalse\@citex[]}}
\def\citelow{\@cghifalse\@ifnextchar [{\@tempswatrue
	\@citex}{\@tempswafalse\@citex[]}}
\def\@cite#1#2{{$\null^{#1}$\if@tempswa\typeout
	{IJCGA warning: optional citation argument 
	ignored: `#2'} \fi}}

\def\pmb#1{\setbox0=\hbox{#1}
	\kern-.025em\copy0\kern-\wd0
	\kern.05em\copy0\kern-\wd0
	\kern-.025em\raise.0433em\box0}

\def\fnt#1#2{\footnotetext{\kern-.3em
	{$^{\mbox{\scriptsize #1}}$}{#2}}}

\def\ps@myheadings{%
    \let\@oddfoot\@empty\let\@evenfoot\@empty
    \def\@evenhead{\slshape\leftmark\hfil}
    \def\@oddhead{\hfil{\slshape\rightmark}}
    \let\@mkboth\@gobbletwo
    \let\sectionmark\@gobble
    \let\subsectionmark\@gobble
    }
\font\tenrm=cmr10
\font\tenit=cmti10 
\font\tenbf=cmbx10
\font\bfit=cmbxti10 at 10pt
\font\ninerm=cmr9
\font\nineit=cmti9
\font\ninebf=cmbx9
\font\eightrm=cmr8

\def\qed{\hbox{${\vcenter{\vbox{			
   \hrule height 0.4pt\hbox{\vrule width 0.4pt height 6pt
   \kern5pt\vrule width 0.4pt}\hrule height 0.4pt}}}$}}

\begin{document}
\setlength{\textheight}{7.7truein}  

\thispagestyle{empty}


\normalsize\textlineskip

\setcounter{page}{1}

\copyrightheading{}	

\vspace*{0.88truein}

\centerline{\bf Quark-Exchange Mechanism of $\gamma d \rightarrow np$
Reaction At 2-6 GeV}
\vspace*{0.4truein}
\centerline{\footnotesize Bruno Julia-Diaz }
\baselineskip=12pt
\centerline{\footnotesize\it Department of Physics, University of Salamanca}
\baselineskip=10pt
\centerline{\footnotesize\it Salamanca, Spain}
\vspace*{12pt}

\centerline{\footnotesize T.-S. H. Lee}
\baselineskip=12pt
\centerline{\footnotesize\it Physics Division, Argonne National Laboratory}
\baselineskip=10pt
\centerline{\footnotesize\it Argonne, Illinois 60439, U.S.A.}
\vspace*{0.228truein}

\publisher{(received date)}{(revised date)}

\vspace*{0.23truein}
\abstract{Within the constituent quark model,
we examine the extent to which the deuteron photo-disintegration at 2-6 GeV
can be described by the quark-exchange mechanism. 
With the parameters constrained by the $np$ scattering, the calculated
differential cross sections disagree with the data in both magnitude and
energy-dependence. The results can be improved if 
we use a smaller size parameter for quark  wavefunctions.
We also find that the on-shell approximation used in a previous investigation 
is not accurate. }


\vspace*{2pt}


\baselineskip=13pt	        
\normalsize              	
\section{Introduction}		
\noindent

It has been well established that nuclear dynamics at low energies can be 
described in terms of hadronic degrees of freedom. Within the
context of Quantum Chromodynamics(QCD),  this hadronic picture
 is expected to break down at sufficiently high energies where the basic 
degrees of freedom must be quarks and gluons. 
It is therefore interesting and 
important to know in which energy region the transition from hadronic
picture to quark-gluon picture takes place. This has been  
the motivation of the experiments on deuteron 
photo-disintegration($\gamma d \rightarrow np$)
conducted at SLAC\cite{1,2} around 1990 and at
Jefferson Laboratory\cite{3,4} in the past few years. More measurements
are expected in the near future.

The limitation of the hadronic picture of deuteron photo-disintegration
at energies above about 1.5 GeV was first suggested in Ref.5.
 The data at 1-6 GeV accumulated at SLAC and JLab
appear\cite{4} to follow
the quark counting rule\cite{6} in the kinematic region where
 the momentum transfer of the reaction becomes larger than
a critical value.  On the other hand, one expects that the non-perturbative
dynamics, such as those due to meson-exchange,
still play an important role in  the transition region.
It is obviously a highly non-trivial problem to interpret these data.
Attempts in this direction have been made in Refs.7-9. 

\begin{figure}[htbp] 
\vspace*{13pt}
\centerline{\psfig{file=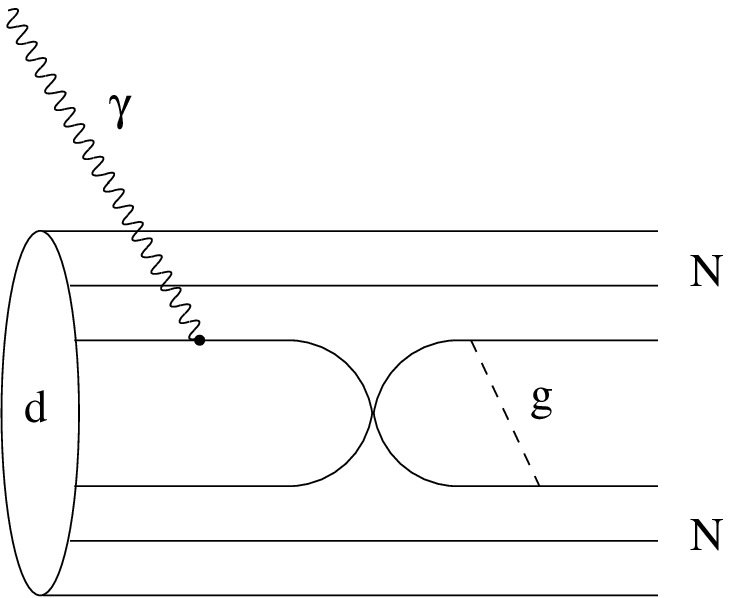}} 
\vspace*{13pt}
\fcaption{The quark-exchange mechanism of $\gamma d \rightarrow np$ reaction.}
\end{figure}
 
In this contribution we would like to report on our recent investigation of
the deuteron photo-disintegration at 2-6 GeV. 
We are motivated by the quark-exchange model of Frankfurt et al.\cite{9}
Based on the quark-exchange mechanism illustrated in Fig.1,
the authors of Ref.9 succeeded in
describing the data at 90 degrees both in
magnitude and energy-dependence. 
On the other hand, their predictions
at other angles failed to describe the data unless a
phenomenological factor is included. 

Because of the complexities of the quark-exchange 
amplitude,  several simplifications were made in the
calculation of Ref.9. First  the on-shell approximation
was used to simplify the loop-integration over the quark propagator. Then
the $\gamma d \rightarrow np$ cross section is cast, with some arguments,
into a product of a factor depending on a convolution over the deuteron
wavefunction and a factor depending on an appropriately evaluated
elastic $np$ cross sections. No explicit calculation in terms of quark-gluon
degrees of freedom is needed. While the simplicity of their final expression
of the differential cross section is very appealing, it is possible that the
difficulties they encountered as well as the successes they achieved
could be due to the  use of these simplifications.
The purpose of this work is to address this question within a model
in which  calculations of the quark-exchange amplitude
can be performed exactly.

\vspace*{2pt}

\section{Quark-Exchange Amplitude}
\noindent
We first observe that apart from the photon-quark interaction vertex, 
the exchange mechanism illustrated in Fig.1 is the same as that 
 studied\cite{fujiwara,bruno} in the
constituent quark model of nucleon-nucleon interaction.
We therefore will perform the calculation within the same theoretical
framework. Thus our approach is not fully relativistic, like all
of the previous studies of multi-quark systems. But this is the only
framework within which an exact calculation of the quark-exchange
amplitude can be performed. 
It is obvious that our intention here is
not a realistic confrontation with the data. Rather it is 
aimed at addressing the theoretical questions mentioned above.

In the center of mass frame, the $\gamma d \rightarrow np$ amplitude
can be written as (suppressing the spin-isospin indices)
\begin{eqnarray}
t(\vec{p},\vec{k}) = <\chi^{(-)}_{\vec{p}}\mid X(\vec{k}) \mid \Phi_d > ,
\end{eqnarray}
where $\vec{k}$ is the incident photon momentum,
$\vec{p}$ the final proton momentum,
$\Phi_d$ the deuteron wavefunction, and 
$\chi^{(-)}_{\vec{p}}$ the final neutron-proton scattering 
wavefunction.
The operator $X(\vec{k})$ represents the quark-exchange mechanism.
The matrix element Eq.(1) can be written as 
\begin{eqnarray}
t(\vec{p},\vec{k}) = \int d\vec{P}_f 
\chi^{(-)*}_{\vec{p}}(\vec{P}_F) T(\vec{P_F},\vec{k}),
\end{eqnarray}
with
\begin{eqnarray}
T(\vec{P}_F,\vec{k}) &=& \int d\vec{P}_1\int d\vec{P}_2
\delta(\vec{P}_1-\vec{P}_F)
\int d\vec{P}^\prime_1 \int d\vec{P}^\prime_2 
\delta(\vec{P}^\prime_1+\vec{P}^\prime_2 + \vec{k}) \nonumber \\
& &X(\vec{P_1},\vec{P_2},\vec{P'_1},\vec{P'_2},\vec{k})  
\Phi_d(\frac{\vec{P}^\prime_1-\vec{P}^\prime_2}{2}).
\end{eqnarray}
where $\vec{P}_1$ and $\vec{P}_2$($\vec{P}^\prime_1$ and $\vec{P}^\prime_2$)
 denote the momenta of the initial(final) two three-quark clusters.
By assuming that the wavefunctions of all three-quark clusters 
in Fig.1 are of Gaussian form ($\phi(r)\sim exp(\frac{-r^2}{2b^2})$ 
in coordinate-space), we can integrate out most of the
quark momenta analytically and obtain
\begin{eqnarray}
X(\vec{P_1},\vec{P_2},\vec{P'_1},\vec{P'_2},\vec{k}) &=& 
\delta({\vec{P_1} + \vec{P_2}-\vec{P'_1}- \vec{P'_2} -\vec{k}\over2})  
{1\over2^3}{1\over2^3} {b^{12} 3^3 \over \pi^6}
 \nonumber \\
&&\int  d\vec{q_{12}}d\vec{q_{56}} \;  
e^{-b^2{q_{12}^2\over 2}} e^{-b^2{q_{56}^2\over 2}} \nonumber \\
&&\int d\vec{q} d\vec{q_p} \; V(q) G(q_p)h(q_p,k) \nonumber \\
&&
exp [ -b^2\left \{ a_{kk} k^2 +a_{kq}{k}\cdot{q}
+a_{kq_p}{k}\cdot{q_p}   +a_{kP_I}{k}\cdot{P_I}
\right. \nonumber \\
&& 
+a_{kP_F}{k}\cdot{P_F}  
+a_{qq}q^2+a_{qq_p}{q}\cdot{q_p}+a_{qP_I}{q}\cdot{P_I} \nonumber \\
&& +a_{qP_F}{q}\cdot{P_F} 
+a_{q_pq_p} q_p^2+a_{q_pP_I}{q_p}\cdot{P_I} \nonumber \\
&&+a_{q_pP_F}{q_p}\cdot{P_F} 
+a_{P_IP_I}P_I^2 +a_{P_IP_F} {P_I}\cdot{P_F} \nonumber \\
&& \left . +a_{P_FP_F}P_F^2   \right \} ] ,
\end{eqnarray}
where $a_{kk}=19/24$, $a_{kq}=-1/2$, $a_{qq}=3/2$, $a_{kq_p}=-5/2$,
$a_{q,q_p}=a_{q_pq_p}=3$, $a_{kP_I}=3/2$, $a_{qP_I}=-1$,
$a_{q_pP_I}=-3$, $a_{P_IP_I}=7/6$, $a_{kP_F}=-1$,
$a_{qP_F}=2$, $a_{q_P,P_I}=3$, $a_{P_IP_F}=-2$, and $a_{P_FP_F}=7/6$.
The parameter $b$ in Eq.(4) defines the range of Gaussian function; 

In Eq.(4), $V(q)$ is a gluon-exchange interaction, $G(q_p)$ is a
quark propagator, and $h(q_p,k) = e_q\bar{u}_{\vec{q}_p}k_\mu \gamma^\mu
 u_{\vec{q}_p-\vec{k}}$ describes the  photon-quark coupling.
 For simplicity, we follow a quark-model  study of $\pi N$ 
scattering\cite{yoshimoto} to set
\begin{eqnarray}
 V(q)=\sum_{i > j}[ -\pi\alpha_s 
(\lambda_i\cdot \lambda_j)(\sigma_i\cdot \sigma_j)
\frac{\Lambda_g^2}{\Lambda^2_g + \vec{q}^2} ],
\end{eqnarray}
where $\lambda_i$ is the color SU(3) generator,
$\alpha_s=1$ and $\Lambda_g=1.087$ GeV were determined from fitting the
$\pi N$ scattering data. The quark propagator is taken to be
\begin{eqnarray}
G(q_p)&=&{1 \over W - \sqrt{\bar{M}_5 +q_p^2} - \sqrt{m_q^2+q_p^2} + 
i\varepsilon} \nonumber \\
&=&{P\over W-\sqrt{\bar{M}_5+q_p^2}-\sqrt{m_q^2+q_p^2}} \nonumber \\
&&-i\pi \delta(W-\sqrt{\bar{M}_5+q_p^2}-\sqrt{m_q^2+q_p^2})
\end{eqnarray}
where $P$ denotes taking 
 the principal-value integration, $m_q=330$ MeV,
 and $\bar{M}_5 = 5 m_q$. 

To carry out the multi-dimensional integration in Eq.(4), we use the
method of Ref. 15 to expand the gluon-exchange interaction and
deuteron wavefunction in terms of Gaussians : 
$V(q) = \sum_i^{N_V} v_i e^{-\alpha_i q^2}$ and
$\Phi_d(q)=\sum_j^{N_G} g_j e^{-\beta_j q^2}$.
We find that very precise expansions can be obtained with $N_V=35$ and
$N_G=50$.
Substituting these expansions into Eqs.(3), and (4), we then obtain
\begin{eqnarray}
T(\vec{P_F},\vec{k})&=&{1\over2^3}{1\over2^3} {b^{12} 3^3 \over \pi^6}
\int  d\vec{q_{12}}d\vec{q_{56}} \;  
e^{-{b^2q_{12}^2\over 2 }} e^{-{b^2q_{56}^2\over 2}} \nonumber \\
&& \sum_i^{N_V} \sum_j^{N_G}\int  d\vec{q_p} \; G(q_p)h(q_p,k)
  v_i g_j  \nonumber \\
&&
\exp(B) \int d\vec{q} d\vec{P_I} \; 
\exp \left (- {\vec{x}\hat{A}\vec{x} \over 2} +\vec{s}\cdot\vec{x} \right)
\label{equpu}
\end{eqnarray}
where $\vec{x}=(\vec{q},\vec{P_I})$ is a two-component vector and
\begin{eqnarray}
B&=& -b^2\left \{a_{kk}k^2 +a_{kq_p}{k}\cdot{q_p} +a_{kP_F}{k}\cdot{P_F}  
+a_{q_pq_p} q_p^2+a_{q_pP_F}  
{q_p}\cdot{P_F} +a_{P_FP_F} P_F^2   \right \},   \nonumber \\
\hat{A} &=&\left (\matrix{
  2 b^2 (a_{qq} +{\alpha_i\over b^2}) & a_{qP_I}  \cr
               a_{qP_I}  & 2 b^2 (a_{P_IP_I}  +{\beta_j\over b^2})    \cr
         } \right ), \nonumber \\
\vec{s}&=& b^2 ( +a_{kq}\vec{k}+a_{qq_p} \vec{q_p}+a_{qP_F}\vec{P_F},
a_{kP_I} \vec{k} +a_{q_pP_I} \vec{q_p}+a_{P_IP_F}  \vec{P_F}). \nonumber
\end{eqnarray}

All integrations in Eq.(7) except that over the propagating quark
 momentum $\vec{q}_p$  can be carried out analytically by using 
the following formula\cite{varga} for a multidimensional gaussian
\begin{eqnarray}
\int d\vec{x} \; e^{-{\vec{x}\hat{A}\vec{x}\over 2}+\vec{s}\cdot\vec{x}} =
\left ( { (2\pi)^N \over det(A)}\right )^{d/2} e^{{1\over2}s \hat{A}^{-1} s}
\end{eqnarray}
where $d$ is the dimension of the space and $N$ is the number of 
variables, $\vec{x}=\{\vec{x_1},\vec{x_2},...,\vec{x_N}\}$ is a N-component
vector.
Schematically, Eq.(7) is then reduced into
\begin{eqnarray}
T(\vec{P}_F,\vec{k}) = 
[ \sum_{i}^{N_V}\sum_{j}^{N_G}v_i, g_j
 \int d \vec{q}_p G(\vec{q}_p)h(q_p,k) 
 I_{i,j}(\vec{P}_F,\vec{k}, \vec{q}_p)] 
\end{eqnarray}
where $ I_{ij}(\vec{P}_F,\vec{k}, \vec{q}_p)$ are from integrating
analytically various
Gaussian functions. We then carry out the remaining integration over 
the propagating quark momentum $\vec{q}_p$ in Eq.(9)
numerically by using the standard substraction method to handle the
singularity of $G(q_p)$ of Eq.(6). 

\begin{figure}[htbp] 
\vspace*{13pt}
\centerline{\psfig{file=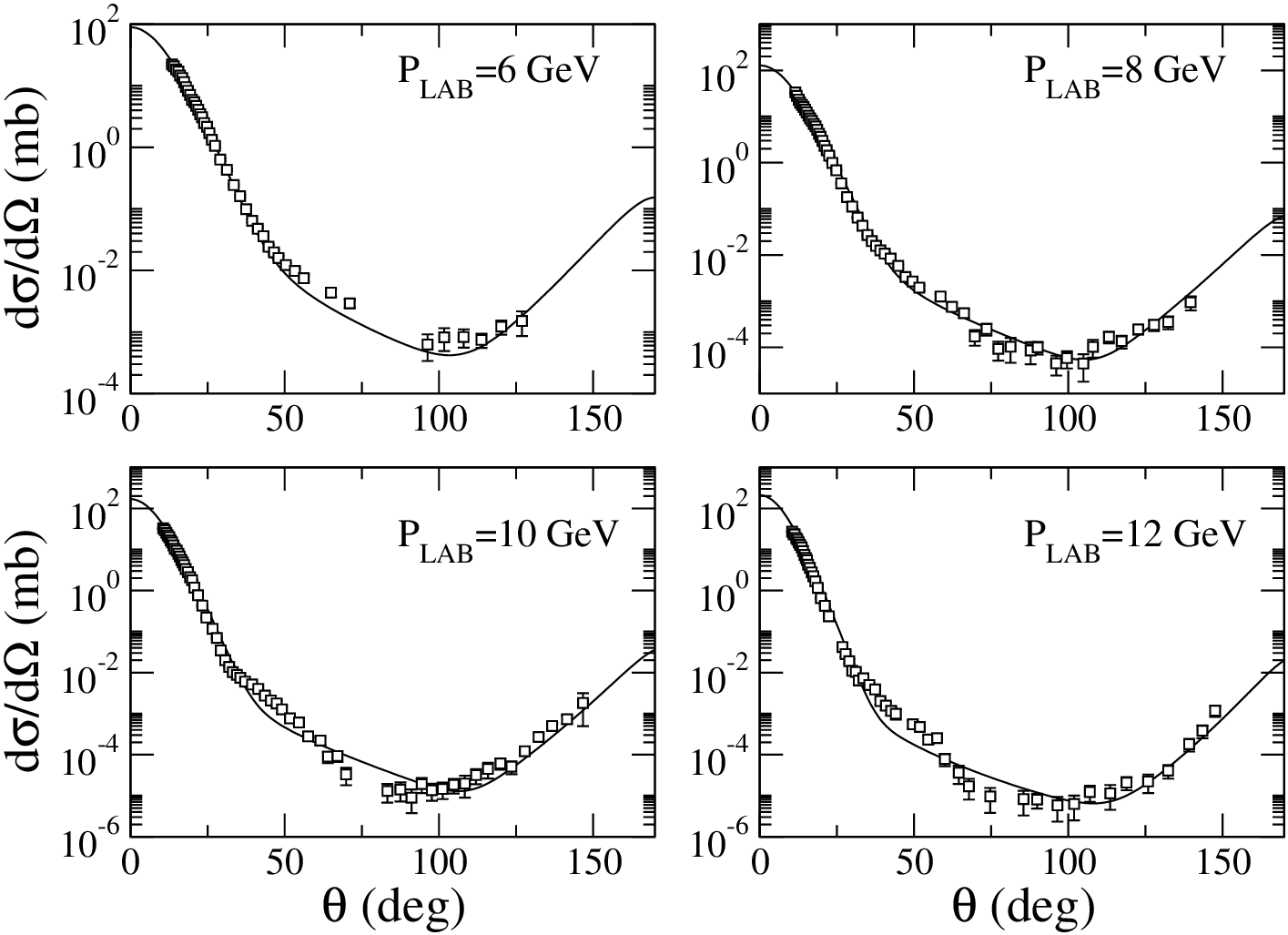,width=8cm}} 
\vspace*{13pt}
\fcaption{Some results from our fits to $np$ elastic scattering 
data.
 The data are from Refs.12-13.}
\end{figure}

\section{Results}
The first step to evaluate the $\gamma d \rightarrow np$ amplitude
 Eq.(2) is to construct a $np$ model for
generating the scattering wavefunction $\chi^{(-)}_{\vec{p}}$.
It can be written in terms of $np$ scattering t-matrix
\begin{eqnarray}
\chi^{(-)*}_{\vec{p}}(p') = \delta(\vec{p}-\vec{p}')
+t(\vec{p},\vec{p}',E)\frac{1}{E - 2 E_N(p')+ i\epsilon}
\end{eqnarray}
where $E=2 E_N(p)$ is the total energy. 
We have found that the $np$ elastic scattering data\cite{10,11} in the
relevant  4-12 GeV region can be
fitted by parameterizing the amplitude $t(\vec{p}^\prime,\vec{p},E)$
as a sum of the  exchanges of $\pi$, $\sigma$, $\rho$,
 $\omega$, and Pomeron. Our fits
 at four energies are displayed  in Fig.2. The details will be given in
Ref.16.

\begin{figure}[htbp] 
\vspace*{13pt}
\centerline{\psfig{file=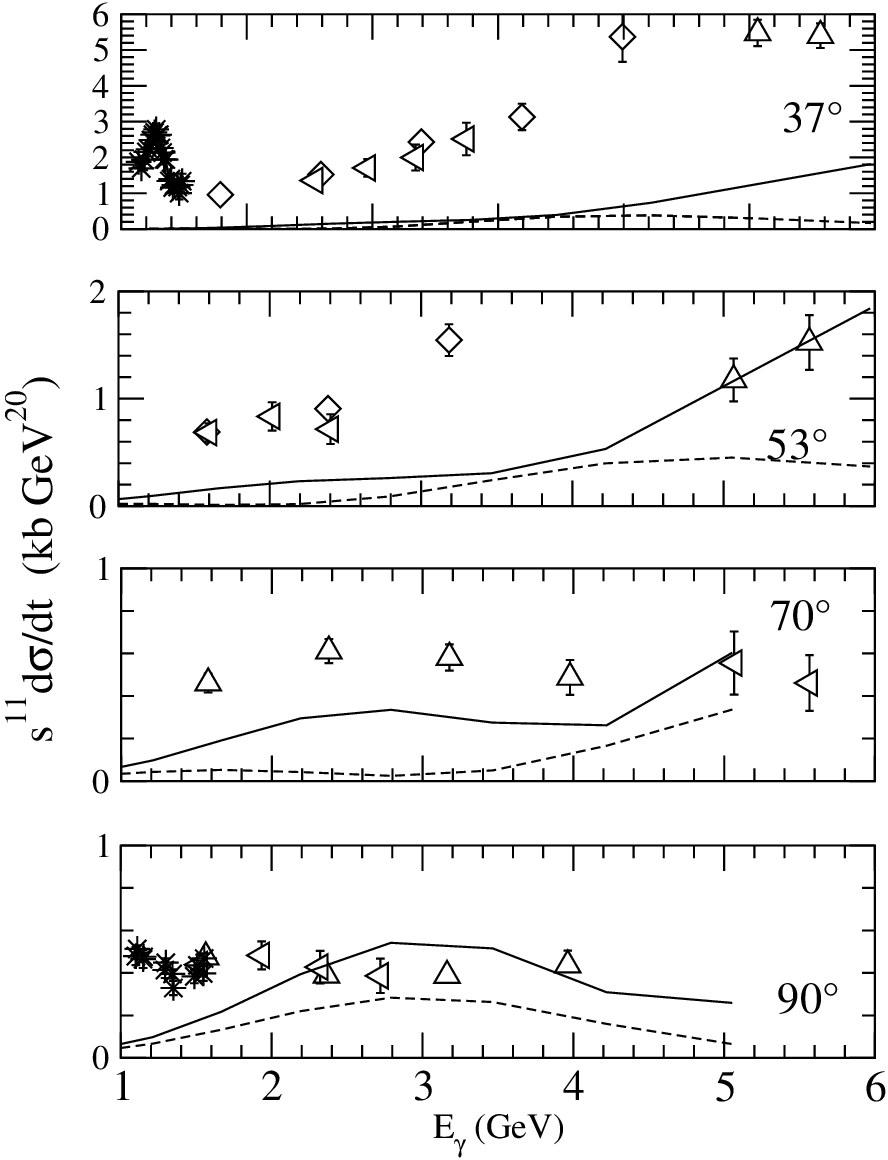,height=9cm}} 
\fcaption{The $\gamma d \rightarrow np$ differential cross sections.
No $np$ final state interaction is included. The solid curves are 
from an exact calculation of Eq.(7), while the dashed curves are from
taking the on-shell approximation used in Ref.9. The data are from Refs.1-4.}
\vspace*{13pt}
\end{figure}

To explore the dynamical content of the quark-exchange amplitude 
$T(\vec{P}_F,\vec{k})$, we first perform
calculations with the final $np$ scattering neglected;
i.e. setting $\chi^{(-)*}_{\vec{p}}(\vec{p}^\prime) = \delta(\vec{p}-
\vec{p}^\prime)$ in evaluating Eq.(2). The deuteron wavefunction is generated 
from the Paris potential.
With the specifications given in section 2, the only free parameter
of the calculations is then
the range parameter $b$ of the quark wavefunctions. We find that
the predicted cross section is very sensitive to this parameter.
If we set $ b = 0.52$ fm used in the quark model of nucleon-nucleon
interaction\cite{bruno}, the predicted cross sections are an order of magnitude
smaller than the data.
This is the first major difficulty of the quark-exchange model considered
here. To get reasonable magnitudes at $90^0$ degrees,
 we need to use a much smaller
value $b=0.25$ fm. The calculated differential cross
 sections are the solid curves shown in Fig.3. 

We now turn to investigating the validity of the on-shell approximation
used in Ref.9. This amounts to performing our calculations by neglecting
the principal-value part of the quark propagator Eq.(6).
The calculated results are the dashed curves in Fig.3.
The difference between the solid and dashed curves indicate very sizable
contributions from the principal-value parts of the integrations.
 Our results strongly suggest that  the
on-shell approximation used in Ref.9 is not accurate.

\begin{figure}[htbp] 
\vspace*{13pt}
\centerline{\psfig{file=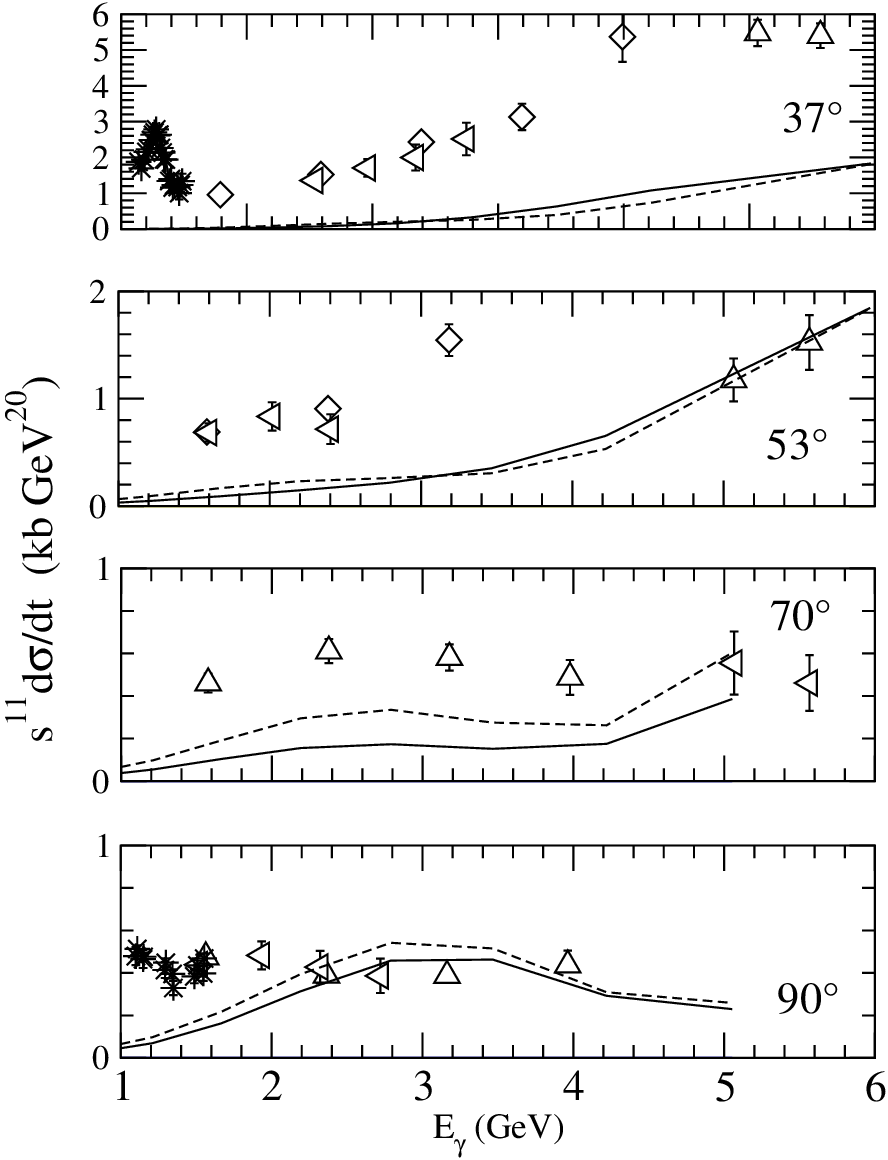,height=9cm}} 
\vspace*{13pt}
\fcaption{The $\gamma d \rightarrow np$ differential cross sections.
The solid curves are from our full calculations, while the dashed curves
are obtained when the final $np$ scattering is neglected. The data
are from Refs.1-4.}
\end{figure}

Our complete calculations are the
solid curves in Fig.4. The dashed curves are obtained when the $np$ final state 
interactions are omitted. We see that the $np$ final state interaction,
constrained by the data shown in Fig.2, do not change our predictions
significantly except at 70 degrees. Clearly our complete calculations
are not able to describe the data. 

To end, we emphasize that this is a very first step toward
a realistic calculation of $\gamma d \rightarrow np$ reactions in 
terms of quark-gluon degrees of freedom. To firmly assess the 
quark-exchange model, several improvements must be made in our
calculations. First, 
we find that our predictions are sensitive to the choice of the
quark mass. This suggests that the use of the
quark propagator Eq.(6) with a 
constituent quark mass 330 MeV may not be correct. We may need to
include the momentum-dependence of quark mass, as suggested by
some QCD model calculations\cite{roberts}. 
Second, the gluon-exchange interaction Eq.(5), taken from
a $\pi N$ study, 
needs to be improved to include more realistic momentum-dependence
and other components of the gluon-exchange interaction. Third, we find that
the main contribution to the cross sections is from the
high momentum component of the deuteron wavefunction in the region where
the employed nonrelativistic Paris potential may not be valid.
It is necessary to generate the deuteron wavefunction from
a model which can also describe $np$ scattering data at high energies,
such as those shown in
Fig.2. It may be possible to obtain such a model by extending
the $\pi NN$ model of Ref.\cite{lee} to higher energies.
Finally, the size parameter $b$ of the three-quark clusters 
 needs to be chosen from a more fundamental consideration.

\nonumsection{Acknowledgments}
\noindent
This work is supported by the U.S. Department of Energy, Nuclear Physics
Division, under Contract No. W-31-109-ENG-38.
B.J. thanks the Ministerio de Ciencia y Tecnologia of Spain
for finantial support, and the hospitality of the Theory Group of 
Argonne National Laboratory.

\end{document}